\documentclass{article}
\usepackage{amssymb}
\usepackage{amsmath}
\usepackage{graphicx}

\begin{document}

\begin{center}
\bigskip {\LARGE \ Ring-down gravitational waves and lensing observables:\
How far can a wormhole mimic those of a black hole?}

\bigskip

\bigskip

Kamal K. Nandi$^{a}$, Ramil N. Izmailov$^{b}$, Almir A. Yanbekov$^{c}$ and
Azat A. Shayakhmetov$^{d}$

\bigskip

Zel'dovich International Center for Astrophysics,

Bashkir State Pedagogical University,

3A, October Revolution Street,

Ufa 450000, RB, Russia

\bigskip

\textbf{Abstract}
\end{center}

It has been argued that the recently detected ring-down gravity waveforms
could be indicative only of the presence of light rings in a horizonless
object, such as a surgical Schwarzschild wormhole, with the frequencies
differing drastically from those of the horizon quasinormal mode frequencies 
$\omega _{\text{QNM}}$ at late times. While the possibility of such a
horizonless alternative is novel by itself, we show by the example of
Ellis-Bronnikov wormhole that the differences in $\omega _{\text{QNM}}$ in
the eikonal limit (large $l$) need not be drastic. This result will be
reached by exploiting the connection between $\omega _{\text{QNM}}$ and the
Bozza strong field lensing parameters. We shall also show that the lensing
observables of the Ellis-Bronnikov wormhole can also be very close to those
of a black hole (say, SgrA$^{\ast }$ hosted by our galaxy) of the same mass.
This situation indicates that the ring-down frequencies and lensing
observables of the Ellis-Bronnikov wormhole can remarkably mimic those of a
black hole. The constraint on wormhole parameter $\gamma $ imposed by
experimental accuracy is briefly discussed. We also provide independent
arguments supporting the stability of the Ellis-Bronnikov wormhole proven
recently.

\begin{center}
---------------------------------------------
\end{center}

\textbf{I. INTRODUCTION}

\baselineskip=4ex%
Direct detection of gravity waves that originated 1.4 billion years ago from
a binary merger is one of the great discoveries of this century [1,2], once
again confirming Einstein's theory of gravity. The detected waves are
assumed to contain the signatures of quasinormal modes (QNM) characteristic
of the formation of a final black hole horizon. Theoretically, these modes
are resonant non-radial deformations induced by external perturbations and
are dictated strictly by the boundary conditions at the horizon, with the
Schwarzschild horizon remaining stable under external perturbations. For the
first time, an alternative source of such waves has been proposed by Cardoso 
\textit{et al. }[3], which is a horizonless, static surgical Schwarzschild
thin-shell wormhole joined at the throat $r_{0}>2M$.

However, the surgical wormhole risks collapse to a point $r_{0}=0$ under
perturbations caused by a moving particle destroying the unstable photon
spheres at $r=3M$. Due to the negative unbound potential of the problem, the
throat would at best be metastable against collapse to $r_{0}=0$ and at
worst, if the joining surface is a classical membrane, be completely
unstable [4]. Granting that the radial test particle motion somehow causes
non-radial deformations of spacetime needed for QNM emission, stability of
the surgical wormhole against such perturbations remains a "completely
uncharted territory" [4].

Stability issues aside, the drastic difference, concluded in [3], in the
fundamental ring-down frequencies between the surgical wormhole and a black
hole of same mass $M$ seems to highlight the topological differences between
a throat and a horizon. We shall exemplify that the difference need not
always be drastic - there could be situations, where wormhole ring-down
modes in the eikonal limit could be very close to those of a black hole of
the same mass. To this end, we note that Jordan frame Brans solutions can
represent wormholes and naked singularities [5], but never black holes, as
has been reported recently by Faraoni \textit{et al.} [6]. We here add that
their conclusion holds true as long as the relevant parameter of the Brans
wormhole solution assumes \textit{real} values as opposed to imaginary ones
(meaning that a throat not topologically changing to a horizon). If the
parameter takes on an \textit{imaginary} value, black hole solution with
vanishing scalar field could result but several arguments in Sec. V indicate
that such an end-state is unlikely to occur in practice. As an example, note
that the Brans II solution can be re-phrased in the Einstein conformal frame
as what is (not widely) known as the horizonless regular Ellis-Bronnikov
wormhole [7,8]. It does not represent a black hole for real values of
parameters but does so for an imaginary value, which is unreachable from a
real regime. Therefore, we should regard the wormhole as an independent
entity by itself that is distinct from a black hole of the same mass.

With regard to the ring-down modes, Konoplya and Zhidenko [9] very recently
studied the dominating low $l$ modes in the gravitational radiation\footnote{%
We thank R.A. Konoplya and A. Zhidenko for pointing out in private correspondence 
the similarity of the problem they dealt with in their paper, especially the 
late-time behavior of the gravitation radiation.}. 
They showed that (i) the $l$ $=2$ $(n=0)$ mode $%
q\omega =1.246-0.192i$ of the Ellis-Bronnikov wormhole has different quality
factor $\sim $ Re($\omega $)/Im($\omega $) from that of the Schwarzschild
black hole for which $M\omega =0.3737-0.0890i$. This means that one can
always differentiate a wormhole from a spherically symmetric black hole in
general relativity, even if the corresponding mass parameters, $q$ and $M$,
are unknown. (ii) The\textit{\ late-time behavior} of ring-down modes for
the $l=2$ axial gravitational perturbations of the Ellis-Bronnikov wormhole
with scalar field mass $q=2.16M$ show the same decay rate as that of a
Schwarzschild black hole of mass $M$ but higher oscillation frequency and
finally (iii) the wormhole, despite the different behaviour of the effective
potential compared to that of the black hole, either rings as a black hole
at all times or rings differently also at all times, depending on the chosen
values of its parameters. In the large $l$ limit, however, it will turn out
that the wormhole and black hole modes of $\omega _{\text{QNM}}$ are almost,
but not exactly, the same. In an earlier work, Konoplya and Zhidenko [10]
developed generic formulas\ for $\omega _{\text{QNM}}$ in the low $l$ limit
for the Morris-Thorne wormhole, static and rotating, using the
Wentzel-Kramers-Brillouin (WKB) method.

The purpose of this paper is to consider the analytic (as opposed to
surgical) horizonless Ellis-Bronnikov wormhole and compare its practically
observable properties with those of a Schwarzschild black hole to see how
far they tally with each other. We shall show that the quantitative
deviations in the large $l$ limit of $\omega _{\text{QNM}}$ and strong field
Bozza lens parameters [11] between the SgrA$^{\ast }$ and Ellis-Bronnikov
wormhole need not be too drastic, indicating that the latter can very well
observationally mimic the black hole. The precision required to distinguish
between the two types of objects imposes a constraint on the wormhole
parameter. Some arguments supporting the recently proven stability of the
Ellis-Bronnikov wormhole are also provided.

We wish to emphasize that we are considering a static compact object merely
as a toy model for SgrA$^{\ast }$ as has been considered, for instance, by
Lacroix and Silk [12], where they commented that a spinning object would be
more appropriate. The reason is that many astrophysical observations of
black holes are not consistent with the static Schwarzschild metric. For
instance, the detection of $106-$ day radio variability in the $\lambda >1$
cm emission from Sgr A$^{\ast }$ signals a small spin ($a\simeq 0.088M$ )
[13] that could lead to new physical phenomenon like superradiance [14].
Additionally, the eikonal limit of Kerr QNMs is not yet fully understood
(see the review in [15]). Given these complications, the results of the
present paper, though limited by the assumption of staticity, could
nevertheless be useful from the heuristic point of view.

In Sec. II, we review Ellis-Bronnikov wormhole including its Schwarzschild
limit. In Sec. II, we quantitatively compute QNM frequencies using strong
field wormhole lensing. In Sec. IV, lensing observables are calculated and a
constraint on the wormhole parameter is obtained. In Sec. V some arguments
supporting the recently proven stability of the Ellis-Bronnikov wormhole.
Sec. VI concludes the paper. We choose units such that $8\pi G=1$, $c=1$
unless specifically restored.

\textbf{II. ELLIS-BRONNIKOV WORMHOLE}

We start with the Einstein field equations that follow from the action with
a minimally coupled scalar field $\phi $. The action and the resulting field
equations are%
\begin{equation}
S=\int \sqrt{-g}d^{4}x\left[ R-\varepsilon g^{\mu \nu }\phi _{,\mu }\phi
_{,\nu }\right]
\end{equation}%
\begin{eqnarray}
R_{\mu \nu } &=&\varepsilon \phi _{,\mu }\phi _{,\nu } \\
\square \phi &\equiv &\phi _{;\mu }^{;\mu }=0,
\end{eqnarray}%
where $\varepsilon $ is a constant, $\phi _{,\mu }$ $\equiv \partial \phi
/\partial x^{\mu }$ and semicolon denotes covariant derivative with respect
to $g_{\mu \nu }$. The source scalar field $\phi $ is assumed to be a ghost
field, defined by $\varepsilon =-1$, that violates all energy conditions.
The Ellis-Bronnikov wormhole solution of Eqs. (2,3) is given by [7,8] 
\begin{eqnarray}
d\tau _{\text{EB}}^{2} &=&Adt^{2}-Bd\ell ^{2}-C(d\theta ^{2}+\sin ^{2}\theta
d\varphi ^{2})], \\
A(\ell ) &=&\exp \left[ -\pi \gamma +2\gamma \tan ^{-1}\left( \frac{\ell }{%
m_{0}}\right) \right] , \\
B(\ell ) &=&A^{-1}(\ell ),C(\ell )=B(\ell )(\ell ^{2}+m_{0}^{2}), \\
\phi (\ell ) &=&\kappa \left[ \frac{\pi }{2}\pm 2\tan ^{-1}\left( \frac{\ell 
}{m_{0}}\right) \right] ,2\kappa ^{2}=1+\gamma ^{2},
\end{eqnarray}%
where $\ell \in (-\infty ,\infty )$, $m_{0}$ and $\gamma $ are arbitrary
constants of integration. This horizonless, traversable, everywhere regular
wormhole for real values of $\gamma >0$ has manifestly two asymptotically
flat regions, one with positive Keplerian mass $M$ ($=m_{0}\gamma $) and the
other with negative mass $-Me^{\pi \gamma }$, situated on either side of a
regular throat at $\ell _{\text{th}}=M$. The throat radius is obtained by
minimizing the areal radius, or from $dC/d\ell =0$. The photon sphere is
defined by the positive root of 
\begin{equation}
\frac{A^{\prime }}{A}=\frac{C^{\prime }}{C},
\end{equation}%
where primes denote differentiation with respect to $\ell $. Thus, the
photon sphere appears at $\ell _{\text{ps}}=2M$. Without loss of rigour, we
henceforth regard $M>0$, together with\ the constant $\gamma >0$, as
independent arbitrary parameters of the solution. Studying circular null
geodesics, Cardoso \textit{et al.} [16] in an earlier work showed that the
QNM frequencies of a black hole in the eikonal limit ($l>>1$), restoring $c$
and retaining their notation, is: 
\begin{eqnarray}
\omega _{\text{QNM}} &=&\Omega _{m}l-i\left( n+\frac{1}{2}\right) \left\vert
\lambda \right\vert , \\
\Omega _{m} &=&c\sqrt{\frac{A_{m}}{C_{m}}},\lambda =c\sqrt{\frac{%
A_{m}C_{m}^{\prime \prime }-A_{m}^{\prime \prime }C_{m}}{2B_{m}C_{m}}},
\end{eqnarray}%
\begin{equation*}
A_{m}\equiv A(\ell _{\text{ps}}),C_{m}\equiv C(\ell _{\text{ps}}),C^{\prime
}\equiv \frac{dC}{d\ell }\text{ etc,}
\end{equation*}%
where $n$ and $l$ , respectively, are the number of overtone and angular
momentum of the perturbation, $\Omega _{m}$ is the angular velocity of the
last circular null geodesic (photon sphere) and $\lambda $ is the Lyapunov
exponent determining the instability time scale. The significance of the
subscript $m$ throughout the paper is that the functions are calculated at
the\textit{\ minimum }radial distance that is the radius of the photon
sphere.

Stefanov \textit{et al. }[17] connected the QNM coefficients in Eq. (9) with
the strong lensing parameters as follows:

\begin{equation}
\Omega _{m}=\frac{c}{u_{m}},\lambda =\frac{c}{u_{m}\overline{a}},
\end{equation}%
where $\overline{a}$ and the minimum impact parameter of the light rays $%
u_{m}$, both appear in the strong field Bozza deflection angle $\alpha
(\theta )$ given by%
\begin{eqnarray}
\alpha (\theta ) &=&-\overline{a}\ln \left( \frac{\theta D_{\text{OL}}}{u_{m}%
}-1\right) +\overline{b},\text{ } \\
\overline{a} &=&\frac{\Omega _{m}}{\lambda },u_{m}=\sqrt{\frac{C_{m}}{A_{m}}}%
,
\end{eqnarray}%
and $\overline{b}$ is another parameter to be found in [11] and calculated
explicitly in the Appendix for the Ellis-Bronnikov wormhole, $D_{\text{OL}}$
is the observer-lens distance, $\theta $ is the independent angular variable
such that $\theta D_{\text{OL}}$ represents the closest approach distance of
light rays. The deflection angle logarithmically diverges when the two
distances, $\theta D_{\text{OL}}$ and $u_{m}$, coincide (meaning photon
capture). It can be verified for the Ellis-Bronnikov wormhole (4)-(7) that

\begin{equation}
\overline{a}=1,
\end{equation}%
\textit{independently} of the values of $M$ and $\gamma $, sharing the same
fundamental property as that of the Schwarzschild black hole. Because of
this remarkable sameness, one would be encouraged to know if the
Ellis-Bronnikov wormhole has a Schwarzschild limit.

\textit{Schwarzschild limit}

It seems little known that the Ellis-Bronnikov wormhole\ (1) reduces
analytically, though by no means trivially, to exact Schwarzschild black
hole. This can be shown rigorously as follows: Identify the constant $%
m_{0}=2B$ in $A(\ell )$ of Eq. (5), transform $\ell \rightarrow r$ by 
\begin{equation}
\ell =r-\frac{B^{2}}{r},
\end{equation}%
where $\ell \in (-\infty ,\infty )$ now maps to $r\in (0,\infty )$. Then one
has $A(\ell )\rightarrow $ $P(r)=\exp \left[ -\pi \gamma +2\gamma \tan
^{-1}\left( \frac{x}{B}\right) \right] $, where $x=\frac{1}{2}\left( r-\frac{%
B^{2}}{r}\right) $. Using the identity $\tan ^{-1}\left( \frac{x}{B}\right)
\equiv 2\tan ^{-1}\left( \frac{x+\sqrt{x^{2}+B^{2}}}{B}\right) -\frac{\pi }{2%
}$, we end up finally with a form of the wormhole solution that happens to
be just the Jordan frame Brans Class II solution [18], rewritten in the
Einstein frame [19]: 
\begin{eqnarray}
d\tau _{\text{EB}}^{2} &\rightarrow &d\tau _{\text{Brans II}%
}^{2}=Pdt^{2}-Qdr^{2}-R\left( d\theta ^{2}+\sin ^{2}\theta d\varphi
^{2}\right) , \\
P(r) &=&\exp \left[ 2\epsilon +4\gamma \tan ^{-1}(r/B)\right] , \\
Q(r) &=&\left( 1+\frac{B^{2}}{r^{2}}\right) ^{2}\exp \left[ 2\zeta -4\gamma
\tan ^{-1}(r/B)\right] , \\
R(r) &=&r^{2}Q(r), \\
\phi (r) &=&\kappa \left[ \pi -2\tan ^{-1}(r/B)\right] ,2\kappa
^{2}=1+\gamma ^{2},
\end{eqnarray}%
where $\epsilon =-\pi \gamma $ and $\zeta =\pi \gamma $ are determined by
the condition of asymptotic flatness.

As a first step, we want to know the extent to which Ellis-Bronnikov
wormhole yields post-post-Newtonian (PPN) Schwarzschild values, so we use
the identity%
\begin{equation}
\tan ^{-1}\left( \frac{r}{B}\right) \equiv \frac{\pi }{2}-\tan ^{-1}\left( 
\frac{B}{r}\right) ,
\end{equation}%
for $r>0$, and identifyting as before the positive mass of one mouth as $M$ (%
$=m_{0}\gamma =2B\gamma $), it can be verified that the metric functions
(17)-(19) admit a Robertson expansion [20] as follows: 
\begin{align}
d\tau _{\text{Brans II}}^{2}& =\left( 1-2\alpha _{1}\frac{M}{r}+2\beta _{1}%
\frac{M^{2}}{r^{2}}-\frac{3}{2}\xi _{1}\frac{M^{3}}{r^{3}}+...\right) dt^{2}
\notag \\
& -\left( 1+2\gamma _{1}\frac{M}{r}+\frac{3\delta _{1}M^{2}}{2r^{2}}+\frac{%
\eta _{1}}{2}\frac{M^{3}}{r^{3}}+...\right) [dr^{2}+r^{2}(d\theta ^{2}+\sin
^{2}\theta d\varphi ^{2})],
\end{align}%
where the PPN\ parameters turn out to be%
\begin{equation}
\alpha _{1}=\beta _{1}=\gamma _{1}=1,\text{ }\delta _{1}=\frac{4}{3}+\frac{1%
}{3\gamma ^{2}},\xi _{1}=\frac{8\gamma ^{2}-1}{9\gamma ^{2}},\eta _{1}=\frac{%
8\gamma ^{2}+5}{3\gamma ^{2}}.
\end{equation}

Since $\alpha _{1}=\beta _{1}=\gamma _{1}=1$, the known \textit{weak field}
tests cannot distinguish between the Ellis-Bronnikov wormhole and the
Schwarzschild black hole as the central gravitating object. To distinguish
them, one would require higher order strong field tests that would in
principle put constraints on $\delta _{1}$, $\xi _{1}$ and $\eta _{1}$.
However, looking at Eqs. (23), it is clear that for real values of $\gamma $%
, there is no way that the parameters may assume the Schwarzschild values $%
\delta _{1}=\xi _{1}=\eta _{1}=1$. The PPN parameters acquire values nearest
to, but not the same as, the Schwarzschild values of unity only when $\gamma
\rightarrow \infty $. To put an observational constraint, say at least on $%
\delta _{1}$, hence on $\gamma $, one could think about measuring the
two-way light deflection $\delta \varphi $ by the Sun up to second order in $%
\left( \frac{M}{b}\right) $, which for the metric (22) works out to (using
the method of Keeton and Petters [21])%
\begin{equation}
\delta \varphi \simeq \frac{4M}{b}+\frac{\pi }{4}\left( 16+\frac{1}{\gamma
^{2}}\right) \left( \frac{M}{b}\right) ^{2},
\end{equation}%
where $b$ is the impact parameter. Unfortunately, due to unsurmountable
technical difficulties, this measurement program has been abandoned [22].
Measuring second order light deflection by the central galactic object is
out of question at this moment. However, it is of interest to note that a
constraint on $\gamma $ can be still imposed from the comparison of the 
\textit{shadows} of the Schwarzschild black hole and Ellis-Bronnikov
wormhole (see Sec. III). Interestingly, the expansion coefficients (23)
suggest that, for the exclusive value $\gamma =-i$, it is possible to obtain
all the Schwarzschild values: $\alpha _{1}=\beta _{1}=\gamma _{1}=\delta
_{1}=\xi _{1}=\eta _{1}=1$.

The next step is to apply on Eqs.(16)-(20) a combination of inversion, Wick
rotation, redefinition of the constant $B$ and an identity, 
\begin{equation}
r=-\frac{B^{2}}{\rho },\text{ }\gamma =-i,\text{ }B=\frac{M}{2\gamma },\tanh
^{-1}(x)\equiv \frac{1}{2}\ln \left( \frac{1+x}{1-x}\right) .
\end{equation}%
The final outcome is the Schwarzschild metric 
\begin{equation}
d\tau _{\text{Brans II}}^{2}\rightarrow d\tau _{\text{Sch}}^{2}=\left( \frac{%
1-\frac{M}{2\rho }}{1+\frac{M}{2\rho }}\right) ^{2}dt^{2}-\left( 1+\frac{M}{%
2\rho }\right) ^{4}\left[ d\rho ^{2}+\rho ^{2}\left( d\theta ^{2}+\sin
^{2}\theta d\varphi ^{2}\right) \right] ,
\end{equation}%
which is what we promised to show.

Returning to the wormhole (16), the radius of the throat and the photon
sphere can be obtained as follows: Using $\ell _{\text{th}}=M$ in Eq. (15),
we have $r_{\text{th}}-\frac{B^{2}}{r_{\text{th}}}=M$, from which it follows
that the throat appears at the isotropic radius 
\begin{equation}
r_{\text{th}}^{\pm }=\frac{M}{2\gamma }\left[ \gamma \pm \sqrt{1+\gamma ^{2}}%
\right] ,
\end{equation}%
but the negative sign has to be discarded as $r_{\text{th}}^{-}$ can become
negative for the wormhole range $\gamma >0$. However, for the black hole
value $\gamma =-i$, the throat has a radius $r_{\text{th}}^{\pm }=\frac{M}{2}
$. Since $r=\frac{M^{2}}{4\rho }$, this radius converts to the Schwarzschild
horizon $r_{\text{th}}^{\pm }\rightarrow \rho _{\text{hor}}=\frac{M}{2}$, as
it should. Similarly, using $\ell _{\text{ps}}=2M$ in Eq. (15), we have $r_{%
\text{ps}}-\frac{B^{2}}{r_{\text{ps}}}=2M$, which yields the isotropic
radius of the photon sphere for the Ellis-Bronnikov wormhole \ 
\begin{equation}
r_{\text{ps}}^{\pm }=\frac{M}{2}\left[ 2\pm \sqrt{4+\frac{1}{\gamma ^{2}}}%
\right] .
\end{equation}%
The sign is to be decided by the physical condition that $r_{\text{ps}}^{\pm
}>r_{\text{th}}^{\pm }=\frac{M}{2}=3.12\times 10^{11}$ cm (for SgrA$^{\ast }$
mass $M=4.22\times 10^{6}M_{\odot }$; see [23]). For the wormhole, the
negative sign has to be discarded as $r_{\text{ps}}^{-}$ can become negative
for $\gamma >0$. Thus, for the wormhole having the mass of SgrA$^{\ast }$,
one has $r_{\text{ps}}^{+\text{EB}}=2M=1.25\times 10^{12}$ cm, obtained at $%
\gamma \rightarrow \infty $. However, for the black hole value $\gamma =-i$,
the photon sphere has a radius $r_{\text{ps}}^{+\text{Sch}}=\rho _{\text{ps}%
}=\frac{M}{2}\left( 2+\sqrt{3}\right) =1.16\times 10^{12}$ cm $>\frac{M}{2}$%
, while $r_{\text{ps}}^{-}$ $<\frac{M}{2}$ is to be discarded. Note that $r_{%
\text{th}}^{\pm }$ can also be obtained directly from the metric (16)\ by
minimizing its areal radius, and similarly, $r_{\text{ps}}^{\pm }$ by
solving Eq. (8).

\textbf{III. STRONG FIELD FIELD LENSING OBSERVABLES}

We shall consider the metric (4)-(7) and the latest observed data for the
supermassive black hole SgrA$^{\ast }$ exemplar, believed to be residing at
the core of our galaxy, for the computation of strong field lensing
observables. The incoming light rays that pass very near to the photon
sphere yield strong field lensing observables. For quantitative comparison,
the most suitable quantity is [11,12] 
\begin{equation}
u_{m}=D_{\text{OL}}\theta _{\infty },
\end{equation}%
where $\theta _{\infty }$ is the observable angular separation between each
set of relativistic images with respect to the central lens. The minimum
impact parameter $u_{m}$, also called the \textit{radius of the shadow} of
the lens, is the central observable to be measured in the currently planned
experiments [12]. As evident from Eqs. (11), the quantitative values of $%
\Omega _{m}$ and $\lambda $ depend solely on the strong lensing observable $%
\overline{a}$, and the minimum impact parameter $u_{m}$, and these
information alone can already distinguish between Schwarzschild and
Ellis-Bronnikov wormhole. Therefore, we consider situations that guarantee $%
u_{m}>\ell _{\text{ps}}=2M$ for lensed images to be possible, that is, when
light is not captured by the photon sphere. We find\ from the second of Eq.
(13) that

\begin{equation}
u_{m}^{\text{EB}}=\sqrt{\frac{C(\ell _{\text{ps}})}{A(\ell _{\text{ps}})}}=M%
\sqrt{\left( 4+\frac{1}{\gamma ^{2}}\right) \exp \left[ 2\pi \gamma -4\gamma
\tan ^{-1}\left( 2\gamma \right) \right] }=D_{\text{OL}}\theta _{\infty }.
\end{equation}%
This equation will be used below to constrain the wormhole parameter $\gamma 
$.

\textit{Constraint on }$\gamma $\textit{:\ Experimental situation}

It is evident from Eq. (30) that , for $\gamma =-i$, we retrieve just the
Schwarzschild value $u_{m}^{\text{Sch}}=3\sqrt{3}M$. Our idea is to
constrain the real values of $\gamma $ in $u_{m}^{\text{EB}}$ in such a
manner that it approaches $u_{m}^{\text{Sch}}$ as closely as possible in
order to support the claim that the Ellis-Bronnikov wormhole could be a
black hole mimicker. In this regard, note that the \textit{lowest} value of $%
u_{m}^{\text{EB}}$ is $2Me$ that appears only at $\gamma \rightarrow \infty $%
. In this case, 
\begin{equation}
\frac{u_{m}^{\text{EB}}}{u_{m}^{\text{Sch}}}=\frac{2e}{3\sqrt{3}}=1.04627.
\end{equation}%
The value of $\gamma $ can be constrained from below by noting that the
angular radius $\theta _{\infty }$ depends only on $\gamma $ once
mass-to-distance ratio, $M/D_{\text{OL}}$, are provided by independent
measurements. Experimental uncertainties in the values of the ratio would
induce uncertainties in $\theta _{\infty }$, which in turn would constrain $%
\gamma $.

Let us look at the current experimental situation focussing on the most
recent work by Johannsen \textit{et al.} [23]. It is to be noted that,
although the realistic situation should involve spin, a final proof of the
Kerr nature of black holes is still lacking [23,24] and worse, unlike the
static case, a regular spinning wormhole reducing to a Kerr black hole in
some limit is a far cry. Further, it has been pointed out in [23] that the
central observable, viz., the angular radius $\theta _{\infty }$ of the
shadow of a Kerr-like solution (that reduces to Kerr black hole when the
deviation parameters are set to zero) is primarily determined by its
mass-to-distance ratio and depends only \textit{weakly} on its spin and
inclination. Relying on this weak dependence on spin, we use the simulated
mass-to-distance ratio for SgrA$^{\ast }$ to constrain the real parameter $%
\gamma $ of the toy Ellis-Bronnikov wormhole, hopefully without committing
much errors. For the Kerr-like metric with an assumed spin parameter $a=0.5M$%
, Johannsen \textit{et al.} [23] combined the seven-station Event Horizon
Telescope (EHT)\footnote{%
The latest EHT, a global network of millimeter-wave Very Long Baseline
Interferometry (VLBI) array, is expected to provide high-angular-resolution
observation of SgrA$^{\ast }$ and M87. The EHT comprises of multiple
different telescopes at multiple different sites all over the world. Because
the EHT telescopes are so far-flung, the effective size of the telescope is
the size of the whole Earth. The shadow of the lens (be it a black or
wormhole) is the main observable target in a direct imaging survey, and this
is what the EHT collaboration aims to observe in the near future, using the
technique of VLBI. The eight observatories comprising EHT are together
capable of directly imaging the shadow of the lens. See the site for
details: http://www.eventhorizontelescope.org/} data at $230$ GHz with
relevant simulations to obtain the SgrA$^{\ast }$ mass $M=4.22\times
10^{6}M_{\odot }$, and its distance $D_{\text{OL}}$($=8.33$ kpc). Using
these values, we get, from the second in Eq. (30)%
\begin{equation}
\theta _{\infty }^{\text{EB}}\left( \gamma \rightarrow \infty \right)
-\theta _{\infty }^{\text{Sch}}\left( \gamma =-i\right) =27.253-26.048=1.205%
\text{ microarcsec.}
\end{equation}%
Since $\theta _{\infty }^{\text{EB}}\left( \gamma \rightarrow \infty \right) 
$ is the lowest value of $\theta _{\infty }^{\text{EB}}\left( \gamma \right) 
$, the exact difference above cannot be reduced further, which signals the
intrinsic difference between the two types of lenses\footnote{%
Illustrative numerologies are as follows: $\theta _{\infty }=\left( u_{m}/D_{%
\text{OL}}\right) \times (206265\times 10^{6})$ microarcsec. For the
Ellis-Bronnikov wormhole of mass $M=4.22\times 10^{6}\times 1.48\times
10^{5} $ cm $=6.245\times 10^{11}$ cm, $u_{m}=2Me=3.395\times 10^{12}$ cm, $%
D_{\text{OL}}=8.33\times 3.085\times 10^{21}=2.569\times 10^{22}$ cm, we
have $\theta _{\infty }^{\text{EB}}=27.253$ microarcsec. For the
Schwarzschild case, $u_{m}=3\sqrt{3}M=3.245\times 10^{12}$ cm, and similarly 
$\theta _{\infty }^{\text{Sch}}=26.048$ microarcsec.}. To find the
constraint on $\gamma $, we plot, for the above $M$, the difference function 
\begin{equation}
\Delta (\gamma )\equiv \theta _{\infty }^{\text{EB}}\left( M,\gamma \right)
-\theta _{\infty }^{\text{Sch}}\left( M,\gamma =-i\right)
\end{equation}%
against $\gamma $ to see for what value of $\gamma $, $\Delta $ becomes
closer to $1.205$ microarcsec. The plot (Fig.1) shows that this happens at $%
\gamma \geq 80$. This is the desired constraint on $\gamma $ from below.

\begin{figure}[tbp]
\includegraphics [width=\linewidth] {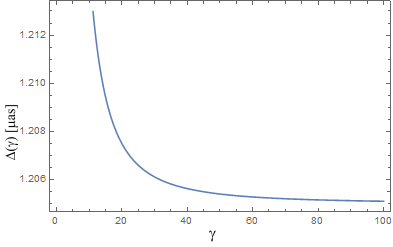}
\caption{Plot of the function $\Delta (\protect\gamma )$ vs the
dimensionless parameter $\protect\gamma $.}
\end{figure}

The question now is whether or not the uncertainty in the current level of
measurement of $\theta _{\infty }$ is smaller than the difference $1.205$
microarcsec just calculated. Once again, based on a reconstructed circular
image of Sgr A$^{\ast }$ from a simulated one-day observing run of the EHT
array, and employing a Markov Chain Monte Carlo (MCMC) algorithm, Johannsen 
\textit{et al.} [23] have demonstrated that such an observation can measure
the angular radius $\theta _{\infty }$ of the shadow of Sgr A$^{\ast }$ to
be $(26.4\pm 1.5)$ microarcsec, i.e., with an uncertainty of $1.5$
microarcsec ($6\%$) and that tight constraints on potential deviations from
the Kerr metric can be obtained\footnote{%
There appears to be a small gap of $0.36$ microarcsec in $\theta _{\infty }$
between the simulated value $26.4$ microarcsec for the Kerr-like case and
the Schwarzschild value $26.04$ microarcsec $\left( a=0\right) $. However,
this gap is expected to be much reduced since the observed Sgr A$^{\ast }$
spin $a\simeq 0.088M$ is far too less than $a=0.5M$, assumed in [25]. The
Kerr lens "identity card" ($\overline{a},\overline{b},u_{m}$) differs little
for $a\approx 0.088M$ from that of the Schwarzschild lens ($a=0$) as the
plots in [26] readily show. Therefore, the value $\theta _{\infty }^{\text{%
Sch}}=26.048$ microarcsec seems quite acceptable at small to moderate spin
values.
\par
\bigskip
\par
{}}. We see that the current uncertainty $1.5$ microarcsec is larger than,
but quite close to, the needed accuracy $\leq 1.205$ microarcsec. Hence, as
of now, measurement of the angular radius of the shadow of our central
galactic object cannot distinguish between the types of lenses but in the
near future it should be possible.

We can also calculate other lensing observables defined in [11] such as the
separation of images $s^{\text{EB}}=\theta _{\infty }$exp$\left[ \frac{1}{%
\overline{a}}(\overline{b}-2\pi )\right] =0.0315$, $s^{\text{Sch}}=0.0321$,
and the ratio\ of fluxes $r=$ exp$\left[ \frac{2\pi }{\overline{a}}\right] $
converted to magnitudes $r_{m}^{\text{EB}}=r_{m}^{\text{Sch}}=2.5\times $ log%
$_{10}\left( r\right) =6.821$. ($\overline{a}=1$ and for $\overline{b}$, see
Appendix).

The zero-mass wormhole ($M=m_{0}\gamma =0$) is obtained by putting $\gamma
=0,m_{0}\neq 0$, which leads to the wormhole metric%
\begin{eqnarray}
d\tau _{\text{EB}}^{2} &=&dt^{2}-d\ell ^{2}-(\ell ^{2}+m_{0}^{2})(d\theta
^{2}+\sin ^{2}\theta d\varphi ^{2})], \\
\phi (\ell ) &=&\frac{1}{\sqrt{2}}\left[ \frac{\pi }{2}\pm 2\tan ^{-1}\left( 
\frac{\ell }{m_{0}}\right) \right] \simeq \text{ const. }\mp \frac{\sqrt{2}%
m_{0}}{\ell }.
\end{eqnarray}%
The constant $m_{0}$ is often interpreted as scalar charge \textit{\`{a} la }%
Wheeleresque\textit{\ }mantra "charge without charge" (see [4,26]). In the
limit $\gamma \rightarrow 0$, the impact parameter $u_{m}^{\text{EB}%
}\rightarrow m_{0}$, which satisfies $u_{m}>\ell _{ps}=0$. Thus, choosing
numerical values for $m_{0}$ equalling the positive mass of Ellis-Bronnikov
wormhole (which is the SgrA$^{\ast }$), we can have exactly the same
observables as in Table I. However, it means that the entire mass of SgrA$%
^{\ast }$ has to be made up of scalar charges without any molecular
structure. This is intriguing but absurd.

\textbf{IV. QNM\ FREQUENCIES}

Since $\overline{a}=1$, we can\textit{\ intuitively} insert the lensing
observable $u_{m}$ in the equation for $\omega _{\text{QNM}}$ derived using
the eikonal limit WKB approximation for Schwarzschild black hole (for
details, see [27]), viz., 
\begin{eqnarray}
\omega _{\text{QNM}} &=&\left( \frac{1}{u_{m}}\right) \left[ \left( l+\frac{1%
}{2}\right) -\frac{1}{3}\left( \frac{5\alpha ^{2}}{12}-\beta +\frac{115}{144}%
\right) l^{-1}+\frac{1}{6}\left( \frac{5\alpha ^{2}}{12}-\beta +\frac{115}{%
144}\right) l^{-2}\right]  \notag \\
&&-i\alpha \left( \frac{1}{u_{m}}\right) \left[ 1+\frac{1}{9}\left( \frac{%
235\alpha ^{2}}{432}+\beta -\frac{1415}{1728}\right) l^{-2}\right] ,
\end{eqnarray}%
where $\alpha \equiv n+\frac{1}{2}$ and $\beta =0,1,-3$ for scalar,
electromagnetic and gravitational perturbations, respectively. We noted that
the original expression for $\omega _{\text{QNM}}$\ derived in (\textit{%
Eq.(3.1)} of [27]) had the same form as above except that, in the
denominator on the left hand side, there was the expression $3\sqrt{3}M$ in
place of $u_{m}$. The surprising thing is that no concept of the radius of
the Schwarzschild black hole shadow $u_{m}$ was used in the WKB method. This
led us to guess that a more generic expression for $\omega _{\text{QNM}}$
should involve $u_{m}$\ in place of the specific Schwarzschild $3\sqrt{3}M$.
The motive for this intuitive generalization is the hope that the frequency
formula (36) be applicable to any spherically symmetric compact object
having a shadow radius $u_{m}$ (including horizonless wormholes) and remains
valid from moderate to large values of $l$. Then one obtains the ratio of
frequencies as:%
\begin{equation}
\frac{\omega _{\text{QNM}}^{\text{Sch}}}{\omega _{\text{QNM}}^{\text{EB}}}=%
\frac{2e}{3\sqrt{3}}=1.04627,
\end{equation}%
which is\textit{\ independent} of $\gamma $, $\beta $, $l$ and $n$. Note
that, for $l\gg 1$, one anyway recovers the eikonal approximation (9) and
the same ratio of frequencies (37) follow. A consequence of our guesswork is
a generic relation from Eqs.(31) and (37), viz., 
\begin{equation}
\omega _{\text{QNM}}^{\text{Sch}}u_{m}^{\text{Sch}}=\omega _{\text{QNM}}^{%
\text{EB}}u_{m}^{\text{EB}}\Rightarrow \omega _{\text{QNM}}u_{m}=\text{%
complex constant,}
\end{equation}%
indicating that the shadows of the lens contain information on $\omega _{%
\text{QNM}}$ and vice versa of any spherically symmetric compact object.

We wish to emphasize that the isotropic form (16) of Ellis-Bronnikov was
derived to show its passage to the Schwarzschild black hole, but it is also
an equally valid coordinate form. We could do all the calculations of
observables using the isotropic form equally well since coordinate choice is
a matter of convenience that does not alter the values of actual
observables. It can be straightforwardly verified using the isotropic metric
(16) that the observables ($\overline{a}$, $\overline{b}$, $u_{m},r,s,\omega
_{\text{QNM}}$) again have exactly the same values. The comparison between
SgrA$^{\ast }$ and the wormhole are summarized at one place in the Table I
below for easy view ($M=4.22\times 10^{6}M_{\odot }$, Schwarzschild case has 
$\gamma =-i$, Ellis-Bronnikov case has $\gamma \geq 80$):

\begin{center}
Table I. Bozza lensing observables for Schwarzschild black hole and
Ellis-Bronnikov wormhole.

\begin{tabular}{|l|l|l|l|l|l|l|l|}
\hline
Lens & $\overline{a}$ & $\overline{b}$ & $u_{m}$ & $r_{m}$ & $s$ & $M\omega
_{\text{QNM}}$($n=0,l=50$) & Re($\omega $)/Im($\omega $) \\ \hline
Sch & $1$ & $-0.4002$ & $5.196$ & $6.821$ & $0.0321$ & $9.718-0.096i$ & $%
101.23$ \\ \hline
EB & $1$ & $-0.4658$ & $5.437$ & $6.821$ & $0.0315$ & $10.168-0.101i$ & $%
100.67$ \\ \hline
\end{tabular}
\end{center}

\textbf{V. STABILITY}

For the wormhole to be an observationally valid alternative to black holes,
the former has to be stable for its very existence. The situation is that,
probably due to the inherent freedom in the choice of perturbation modes,
there have been many differing claims in the literature, of which some are
mentioned here. Previously, Armend\'{a}riz-Pic\'{o}n [28] showed that
massless Ellis-Bronnikov wormhole and at least a non-zero measure set of
massive Ellis-Bronnikov wormholes are stable. But it is subsequently argued
by Gonz\'{a}lez \textit{et al. }[29,30] that the linear stability analysis
in [28] applies only to a restricted class of perturbations, that requires
the perturbed scalar field to vanish at the throat, $\delta \phi (\ell _{%
\text{th}})=0$. Using numerical simulations, they conclude that the wormhole
is unstable under both linear and non-linear perturbations\ such that it
either expands away to infinity or collapse into Schwarzschild black hole.
The conclusion of instability for the phantom scalar field has been
supported also by Bronnikov \textit{et al. }[31].

Below, we wish to point out that, while the emergence of an apparent horizon
in the simulation in [30] is an interesting result based on the particular
mode of perturbation, the conclusion of collapse to black hole seems
arguable for the following reasons:

First, Gonz\'{a}lez \textit{et al. }[30] take the appearance\ of\ apparent
horizon to be a "strong indication" for the formation of an event horizon at
a later stage of collapse. Such a hope might be belied since, as they too
noted, the apparent horizon is both foliation and observer dependent notion
[32]. The main thing is that, its existence is not even mandatory for the
event horizon. It is quite possible to foliate the Schwarzschild geometry in
such a way that there is never any apparent horizon, despite the fact that
there is certainly an event horizon [33].

Second, a more recent\ stability analysis by Novikov and Shatskiy [34] show
that the zero mass wormhole, with the stress decomposed in a clever way, is
stable under spherical perturbations (no collapse, no expansion). The stress
structure being exactly the same for massive Ellis-Bronnikov wormhole as well%
\footnote{%
The stress tensor threading the massive Ellis-Bronnikov wormhole has the
same decomposable components $\rho =-\frac{m^{2}(1+\gamma ^{2})}{\left( \ell
^{2}+m^{2}\right) ^{2}}\exp \left[ -\gamma \left\{ \pi -2\tan ^{-1}\left( 
\frac{\ell }{m}\right) \right\} \right] $, $p_{r}=\rho $, $p_{\theta
}=p_{\varphi }=-\rho .$ Both the Weak Energy Condition (WEC), $\rho \geq 0$
and the Null Energy Condition (NEC), $\rho +p_{r}\geq 0$ are violated. For $%
\gamma =0$, one has the stress of the zero mass case decomposed by Novikov
and Shatskiy [30].}, the same analysis can be extended to this case.
However, there is a simpler, and mathematical, argument: Note that only the
exclusive value of the parameter $\gamma =-i$ in metric (16) yields the
exact Schwarzschild black hole, with $\phi =0$ as was shown. If the
wormhole, for which $\gamma $ must always be \textit{real}, has to really
collapse to a black hole, the parameter $\gamma $ has to jump from real line
into a point on the complex line, augering a sudden topology change! This is
absurd, since topology change is against normal experience, at least, on a
macroscopic scale [35]. A very recent work by Faraoni \textit{et al.} [6]
concludes that Brans solutions cannot collapse into black holes.
Ellis-Bronnikov wormhole is just such a solution being the Einstein frame
variant of the Brans II solution [18], and the same conclusion holds.

Third, we should note that ring-down gravity waves are generated by
non-spherical deformations induced by external perturbations. Meanwhile,
Bronnikov and Rubin [36] argue that the non-spherical perturbation modes
must probably be more stable than the spherical ones, since the effective
potential for the perturbations contains centrifugal (and other higher
multipoles) barriers, like in the Regge-Wheeler or Zerilli potentials. In
fact, stability under non-spherical perturbation is indirectly supported by
the negative imaginary part $\omega _{I}$ of the QNM modes as argued in
[15,27]. Eq. (11) with a positive $\overline{a}$ guarantees that $\lambda >0$
or $\omega _{I}<0$. By the same token, a precise observation of QNM modes
would also constitute a test for the existence or otherwise of scalar hair $%
\phi $ in the wormhole [15,37].

There exists yet another entirely different window to look at the stability
issue, viz., via Tangherlini's approach\ [38] of "non-deterministic,
pre-quantal statistical simulation" of photon motion in a medium yielding
reflection ($R$) and transmission ($T$) coefficients across a surface in the
medium. Taking into account the generic feature in curved space-time,
namely, that observations depend on the location of the observer, this
approach yields observer-dependent \textit{perception of stability} of the
wormhole in terms of these coefficients. While one observer perceives
instability, another observer might perceive stability (see, for details,
[39]).

\textbf{VI. CONCLUSIONS}

Most of the numerous works on QNM frequencies beginning with the seminal
work by Vishveshwara [40] in 1970 until its most recent application to
wormholes [3,9] involve spherically symmetric static sources as toy models.
However, it is to be remarked that spin is an important factor and many
astrophysical observations of black holes are inconsistent with the
Schwarzschild metric\footnote{%
We thank an anonymous referee for directing our attention to this point, as
well as to the question of how $\gamma $ could be constrained by
observations.}. A glimpse of such inconsistency is provided by the observed
radio variability in the emission spectrum from SgrA$^{\ast }$ believed to
be induced by a small spin of an assumed Kerr black hole [13]. Nonetheless,
studies using static sources provide very useful information on the mode
spectrum including the more interesting case of low-lying frequencies ($%
n=0,l=2$ is the dominant gravitational wave mode giving the famous $M\omega
_{\text{QNM}}^{\text{Sch}}=0.3737-0.0890i$).

Cardoso \textit{et al. }[3] consider a static surgical wormhole (born out of
cut-paste surgery joining two copies of Schwarzschild black holes) as a
heuristic model that could be extended by including other effects such as
spin but, they argue, none of these effects is expected to change the
qualitative picture. The present work has to be regarded as an improvement
on the same idea, where the artificial surgical wormhole has been replaced
by the regular stable Ellis-Bronnikov wormhole. The latter, being a solution
of general relativity with a well defined source, stands a better chance for
its occurrence in nature as a competing astrophysical object. This
notwithstanding, the results of the present paper should be taken as
indicative rather than concrete due to lack of spin. An exact treatment
incorporating spin would require a separate follow-up investigation to see
if the frequencies emanated from a spinning black hole can be connected with
its strong field lensing observables in a manner discovered by Stefanov 
\textit{et al. }[17] for the static case. It is premature to say if this
connection exists at all but at least the strong field observables for SgrA$%
^{\ast }$ with its estimated small spin ($a\simeq 0.088M$) in the Kerr
metric do not differ appreciably from their static values ($a=0$), as the
plots of the lens "identity card" ($\overline{a}$, $\overline{b}$, $u_{m}$)
in [25] show.

The merit of the chosen Ellis-Bronnikov wormhole is that it is
observationally indistinguishable from the Schwarzschild black hole in the
weak field regime since the PPN parameters are the same ($\alpha _{1}=\beta
_{1}=\gamma _{1}=1$). This result raises the possibility if this wormhole
can act as a black hole mimicker beyond PPN approximation. We argue that it
can, but only within the experimental accuracy as available today. A better
accuracy in the future will certainly distinguish between the two objects.
Black hole mimickers are not unknown in the literature. For instance,
gravastar models mimicking black holes have been investigated by Chirenti
and Rezzolla [41,42]. Once again, unlike artificial gravastars, the
Ellis-Bronnikov wormhole is more natural and much simpler. Moreover, it has
been shown that the wormhole reduces exactly to Schwarzschild black hole
under the special choice $\gamma =-i$. One would then think that the
Ellis-Bronnikov wormhole for different values of \textit{real} $\gamma $
would lead to observable signatures very different from those of
Schwarzschild black hole obtained at \textit{imaginary} $\gamma $.
Remarkably, this need \textit{not }be the case! It was shown that the main
observable $u_{m}$ rapidly saturates to $2Me$ at $\gamma \rightarrow \infty $%
, which is indeed not much different from the Schwarzschild value $3\sqrt{3}%
M $. In this sense, Ellis-Bronnikov wormhole can be regarded as assuming 
\textit{an eternal identity} by itself, just like the classical
Schwarzschild black hole.

We applied our calculations to the object residing at the center of our
galaxy that is speculated to be a black hole (SgrA$^{\ast }$) of mass $%
M=4.22\times 10^{6}M_{\odot }$ situated at a distance $D_{\text{OL}}=8.33$
kpc [23]. If instead we regard the object as Ellis-Bronnikov wormhole, then
it turns out that both the objects remarkably share the same value of Bozza
strong field lensing parameter, $\overline{a}=1$. It was further shown that
the ratio between the black and wormhole of the same mass with regard to
ring-down gravitational wave mode in the eikonal limit is set by $\frac{%
\omega _{\text{QNM}}^{\text{Sch}}}{\omega _{\text{QNM}}^{\text{EB}}}=\frac{2e%
}{3\sqrt{3}}=1.04627$ independently of $M,\gamma $, $l$ and $n$. This ratio
cannot be reduced further as the object either rings as a black hole at all
times or rings differently also at all times, depending on the chosen values
of its parameters [9]. It was also calculated that $\theta _{\infty }^{\text{%
EB}}=27.253$ microarcsec, $\theta _{\infty }^{\text{Sch}}=26.048$
microarcsec, which differ just by $1.205$ microarcsec. Other specified
observables [11] were also calculated such as the separation of relativistic
images $s^{\text{EB}}=\theta _{\infty }^{\text{EB}}$exp$\left[ \frac{1}{%
\overline{a}}(\overline{b}-2\pi )\right] =0.0315$, $s^{\text{Sch}}=0.0321$,
the ratio\ of fluxes $r=$ exp$\left[ \frac{2\pi }{\overline{a}}\right] $
converted to magnitudes yields $r_{m}^{\text{EB}}=r_{m}^{\text{Sch}%
}=2.5\times $ log$_{10}\left( r\right) =6.821$, which intriguingly is yet
another exact equality due to $\overline{a}=1$. All the obtained results are
summarized in Table I for easy comparison. It is evident that the
observables for the black and wormhole are quite close, and some are exactly
the same, giving strength to the idea that the Ellis-Bronnikov wormhole can
act as a black hole mimicker within experimental accuracy.

This raises a very relevant question about the main observable and the
current experimental accuracy, and how it can constrain $\gamma $. The
central observable is the angular radius $\theta _{\infty }$ of the shadow
of the object, which is primarily determined by its mass-to-distance ratio
with a weak dependence on spin within the Kerr metric of the theory of
general relativity. If the theory is violated, the shadow size may also
depend strongly on parametric deviations from the Kerr metric. The result
and uncertainty in $\theta _{\infty }$ from a simulated one-day observing
run of the seven station EHT demonstrate that such an observation can
measure $\theta _{\infty }$ of Sgr A$^{\ast }$ with an uncertainty of $1.5$
microarcsec $(6\%)$[23]. (The possibility of directly imaging the shadow of
the lens in the not-too-distant future is quite promising [12,43]). We
calculated in Eq.(32) that the level of accuracy needed to distinguish
between the Schwarzschild black hole and the Ellis-Bronnikov wormhole of the
same mass and distance is $1.205$ microarcsec. The plot of the difference
function $\Delta (\gamma )$ of Eq. (33) then shows that the constraint is $%
\gamma \geq 80$ (Fig.1).

A final remark: Despite intriguingly similar, even the same, observable
values, it is our conviction that the Ellis-Bronnikov wormhole for real
values of $\gamma $ would survive as a topological object of its own class,
remaining fundamentally distinct from a Schwarzschild black hole. This would
be expected because a real $\gamma >0$ cannot jump to $\gamma =-i$, augering
a spontaneous topology change against experience [4,36]. By an intuitive
extension, it is tempting to elevate this conviction into a principle: 
\textit{Collapse of any object will lead to a final state definable only
within the parameter range of the initial object and not to a state defined
by parameters outside that range.}

\textbf{Acknowledgment}

Part of the work was supported by the Russian Foundation for Basic Research
(RFBR) under Grant No.16-32-00323. We thank two anonymous referees for their
insightful comments that helped us improve the paper.

\textbf{References}

[1] B. P. Abbott \textit{et al. }(LIGO/Virgo Scientific Collaboration),
Phys. Rev. Lett. \textbf{116}, 061102 (2016).

[2] B. P. Abbott \textit{et al. (}The LIGO/Virgo Scientific Collaboration),
Phys. Rev. Lett. \textbf{116}, 061101 (2016).

[3] V. Cardoso, E. Franzin and P. Pani, Phys. Rev. Lett. \textbf{116},
171101 (2016); 089902(E)(2016).

[4] M. Visser, \textit{Lorentzian Wormholes }$-$ \textit{From Einstein To
Hawking} (AIP, New York, 1996).

[5] K.K. Nandi, B. Bhattacharjee, S.M.K. Alam and J. Evans, Phys. Rev. D 
\textbf{57}, 823 (1998).

[6] V. Faraoni, F. Hammad and S.D. Belknap-Keet, Phys. Rev. D \textbf{94},
104019 (2016).

[7] H.G. Ellis, J. Math. Phys.\textbf{\ (}N.Y.\textbf{)14}, 104 (1973); 
\textbf{15}, 520(E) (1974).

[8] K.A. Bronnikov, Acta Phys. Pol. B\textbf{\ 4}, 251 (1973).

[9] R.A. Konoplya and A. Zhidenko, J. Cosmol. Astropart. Phys. 12 (2016) 043.

[10] R.A. Konoplya and A. Zhidenko, Phys. Rev. D \textbf{81}, 124036 (2010).

[11] V. Bozza, Phys. Rev. D \textbf{66}, 103001 (2002).

[12] T. Lacroix and J. Silk, Astron. Astrophys. \textbf{554}, A 36 (2013).

[13] S. Liu and F. Melia, Astrophys. J. \textbf{573}, L23 (2002).

[14] K. Kokkotas and B. Schmidt, Living Rev. Relativity, \textbf{2, }2
(1999).

[15] E. Berti, V. Cardoso and A.O. Starinets, Classical Quantum Gravity 
\textbf{26}, 163001 (2009).

[16] V. Cardoso, A. S. Miranda, E. Berti, H. Witek and V. T. Zanchin, Phys.
Rev. D \textbf{79}, 064016 (2009).

[17] I. Zh. Stefanov, S. S. Yazadjiev and G. G. Gyulchev, Phys. Rev. Lett. 
\textbf{104}, 251103 (2010).

[18] K.K. Nandi, A. Islam and J. Evans, Phys. Rev. D \textbf{55}, 2497
(1997).

[19] K. K. Nandi and Y.-Z. Zhang, Phys. Rev. D \textbf{70}, 044040 (2004).

[20] S. Weinberg, \textit{Gravitation and Cosmology} (John Wiley \& Sons,
New York,1972).

[21] C. R. Keeton and A. O. Petters, Phys. Rev. D \textbf{72},104006 (2005).

[22] J. Bodenner and C.M. Will, Am. J. Phys. \textbf{71}, 770 (2003).

[23] T. Johannsen, C.Wang, A.E. Broderick, S.S. Doeleman, V.L. Fish, A.
Loeb, D. Psaltis, Phys. Rev. Lett. \textbf{117}, 091101 (2016).

[24] D. Psaltis, Living Rev. Relativity \textbf{11}, 9 (2008).

[25] V. Bozza, Phys. Rev. D \textbf{67}, 103006 (2003).

[26] K. K. Nandi, Y.-Z. Zhang and A. V. Zakharov, Phys. Rev. D \textbf{74},
024020 (2006).

[27] S. Iyer, Phys. Rev. D\textbf{\ 35}, 3632 (1987).

[28] C. Armend\'{a}riz-Pic\'{o}n, Phys. Rev. D \textbf{65}, 104010 (2002).

[29] J. A. Gonz\'{a}lez, F.S. Guzm\'{a}n and O. Sarbach, Classical Quantum
Gravity \textbf{26}, 015010 (2009).

[30] J. A. Gonz\'{a}lez, F.S. Guzm\'{a}n and O. Sarbach, Class. Quantum
Grav. \textbf{26}, 015011 (2009).

[31] K. A. Bronnikov, R. A. Konoplya and A. Zhidenko, Phys. Rev. D \textbf{86}, 024028 (2012).

[32] I. Booth, Can. J. Phys. \textbf{83}, 1073 (2005).

[33] R. M. Wald and V. Iyer. Phys. Rev. D \textbf{44}, R3719 (1991).

[34] I. D. Novikov and A. A. Shatskiy, J. Exp. Theor. Phys. \textbf{114},
801 (2012).

[35] S. W. Hawking, Phys. Rev. D\textbf{\ 37}, 904 (1988).

[36] K. A. Bronnikov and S. G. Rubin, \textit{Lectures on Gravitation and
Cosmology }(Moscow Engineering Physics Institute, Moscow, 2008).

[37] R. A. Konoplya and A. Zhidenko, Rev. Mod. Phys. \textbf{83}, 793 (2011).

[38] F. R. Tangherlini, Phys. Rev. A \textbf{12}, 139 (1975).

[39] K. K. Nandi, A. A. Potapov, R. N. Izmailov, A.Tamang and J. C. Evans,
Phys. Rev. D \textbf{93}, 104044 (2016).

[40] C.V. Vishveshwara, Nature (London) \textbf{227}, 936 (1970).

[41] C.B.M.H. Chirenti and L. Rezzolla, Classical Quantum Gravity \textbf{24}%
, 4191 (2007).

[42] C.B.M.H. Chirenti and L. Rezzolla, Phys. Rev. D \textbf{94}, 084016
(2016).

[43] A. Ricarte and J. Dexter, Mon. Not. R. Astron. Soc. \textbf{446}, 1973
(2015).

\begin{center}
\textbf{APPENDIX: CALCULATION OF }$\overline{b}$\textbf{\ FOR
ELLIS-BRONNIKOV WORMHOLE }
\end{center}

In the expression for $\alpha (\theta )$, $\theta $ is an independent
angular variable designating different rays and since $\overline{a}=1$, the
only quantities that can be expressed in terms of generic mass $M$ are the
minimum impact parameter $u_{m}$ and $\overline{b}$. The $u_{m}$ has already
been expressed as such in Eq. (30). To obtain the functional expression for $%
\overline{b}$ in terms of $M$, it is necessary to briefly state its origin
as developed by Bozza [11]. Thus the Ellis-Bronnikov metric is taken as [see
Eq. (4)]

\begin{equation*}
d\tau ^{2}=A(\ell )dt^{2}-B(\ell )dr^{2}-C(\ell )\left( d\theta ^{2}+\sin
^{2}\theta d\varphi ^{2}\right) .
\end{equation*}

A photon incoming from infinity with arbitrary impact parameter $u$, will be
deviated while approaching the black hole. The light ray will reach a
closest approach distance $\ell _{0}$ and then emerge in another direction.
The two distances are generically related by%
\begin{equation*}
u=\sqrt{\frac{C(\ell _{0})}{A(\ell _{0})}}.
\end{equation*}%
The minimum impact parameter is 
\begin{equation*}
u_{m}=\sqrt{\frac{C(\ell _{m})}{A(\ell _{m})}},
\end{equation*}%
We shall be using the Ellis-Bronnikov metric functions (4) that yield%
\begin{equation*}
\ell _{m}=\ell _{\text{ps}}=2M=2m_{0}\gamma .
\end{equation*}%
The deflection angle

\begin{equation*}
\alpha (\theta )=-\overline{a}\ln \left( \frac{\theta D_{\text{OL}}}{u_{m}}%
-1\right) +\overline{b}
\end{equation*}%
can be expressed in a mass-dependent form. We just cite here the expression
(Eq. (37) from [11]): 
\begin{eqnarray*}
\overline{b} &=&-\pi +b_{R}+\overline{a}\log \left( \frac{2\beta _{m}}{y_{m}}%
\right) ,\text{ }y_{m}=A(\ell _{m}) \\
b_{R} &=&I_{R}(\ell _{m})=\int_{0}^{1}g(z,\ell _{m})dz\text{, }
\end{eqnarray*}%
and $\beta _{m}$ is an expression involving derivatives of metric functions
(see Eq.(24) of [11]). Omitting the detailed generic expressions for $\beta
_{m}$ and $g(z,\ell _{m})$, we only report here the final expressions for $%
b_{R}$ and $\frac{2\beta _{m}}{y_{m}}$ for the Ellis-Bronnikov wormhole in
terms of $M$. The integrand $g(z,\ell _{m})$ has a formidable expression,
that has been calculated and numerically integrated by using \textit{%
Mathematica 9.1}.\ To explicitly show the mass dependence of $\overline{b}$,
define%
\begin{equation*}
K_{1}=\frac{\exp (-\pi \gamma )}{M}\left[ \exp (\pi \gamma )-\exp \left\{
2\gamma \tan ^{-1}(2M)\right\} \right] \sqrt{(1+4M^{2})/K_{2}}
\end{equation*}%
\begin{eqnarray*}
K_{2} &=&\exp [-\pi \gamma +2\gamma \tan ^{-1}(2M)] \\
K_{3} &=&\exp \left[ 2\gamma \left\{ \tan ^{-1}(2M)+\tan ^{-1}\cot \left( 
\frac{\log K_{2}}{2\gamma }\right) \right\} \right]  \\
K_{4} &=&\ln \left[ \exp (-\pi \gamma )\left\{ (1-z)\exp [2\gamma \tan
^{-1}(2M)]+z\exp (\pi \gamma )\right\} \right]  \\
K_{5} &=&\frac{z}{\gamma ^{2}}\left[ \exp (-\pi \gamma )-\exp \left\{
-2\gamma \right\} \tan ^{-1}(2M)\right]  \\
K_{6} &=&z[4M^{2}-1-12M\gamma +4\gamma ^{2}]\exp (\pi \gamma ) \\
&&+\left[ 8\gamma \left( \gamma -M\right) +z\left( 1-4M^{2}+12M\gamma
-4\gamma ^{2}\right) \right] \exp \left\{ 2\gamma \tan ^{-1}(2M)\right\}  \\
K_{7} &=&\exp \left[ 2\gamma \tan ^{-1}\cot \left( \frac{\log K_{4}}{2\gamma 
}\right) \right]  \\
K_{8} &=&(1-z)\exp \left[ 2\gamma \tan ^{-1}(2M)\right] +z\exp (\pi \gamma )
\\
K_{9} &=&\exp \left[ -\pi \gamma -2\gamma \tan ^{-1}(2M)-2\gamma \tan
^{-1}\cot \left( \frac{\log K_{4}}{2\gamma }\right) \right]  \\
K_{10} &=&(1+4M^{2})\left[ (z-1)\exp \left\{ 2\gamma \tan ^{-1}(2M)\right\}
-z\exp (\pi \gamma )\right] \times \sin \left( \frac{K_{4}}{2\gamma }\right)
,
\end{eqnarray*}

Then

\begin{equation}
g(z,\ell _{m})\equiv g(z,M,\gamma )=K_{1}\times \left( \frac{-2\sqrt{K_{3}}}{%
\sqrt{K_{5}K_{6}}}+\frac{\sqrt{K_{7}K_{8}}}{\sqrt{K_{2}+K_{9}K_{10}}}\right)
.  \tag{A1}
\end{equation}%
We have verified that the numerical integration 
\begin{equation}
b_{R}=\int_{0}^{1}g(z,M,\gamma )dz  \tag{A2}
\end{equation}%
does yield the Schwarzschild value (for $\gamma =-i$), $b_{R}=0.9496$.
Furthermore, it can be verified that 
\begin{equation}
\log \left( \frac{2\beta _{m}}{y_{m}}\right) =\log \left[ \frac{\exp \left[
-4\gamma \tan ^{-1}(2\gamma )\right] \left[ \exp (\pi \gamma )-\exp \left\{
2\gamma \tan ^{-1}(2\gamma )\right\} \right] ^{2}\left[ 1+4\gamma ^{2}\right]
}{2\gamma ^{2}}\right]  \tag{A3}
\end{equation}%
This yields the exact Schwarzschild value (for $\gamma =-i$), $\log \left( 
\frac{2\beta _{m}}{y_{m}}\right) =\log 6=1.7917$. Collecting the results,
one has $\overline{b}^{\text{Sch}}=-\pi +b_{R}+\overline{a}\log \left( \frac{%
2\beta _{m}}{y_{m}}\right) =-0.4002$, just as in [11]. For the
Ellis-Bronnikov wormhole, it was noted that the observable values rapidly
saturate at $\gamma \gtrsim 80$, so at large real $\gamma $, it can be
verified that, for the same mass as that of the black hole,

\begin{equation}
b_{R}=0.8999,  \notag
\end{equation}

\begin{equation}
\log \left( \frac{2\beta _{m}}{y_{m}}\right) =\log \left( 5.905\right)
=1.7758  \notag
\end{equation}

\begin{equation}
\overline{b}^{\text{EB}}=-\pi+b_{R}+\overline{a}\log\left(\frac{2\beta_{m}}{%
y_{m}}\right) =-0.4658  \tag{A4}
\end{equation}

This value\ $\overline{b}^{\text{EB}}$ was used in the text to calculate the
separation of images $s^{\text{EB}}=0.0315$. Note that $b_{R}$ is a result
of a definite integral (A2) giving definite numerical values for black and
wormhole cases, so $\overline{b}$ is independent of coordinate choices in
each case.

\bigskip

\bigskip

\begin{center}
\ 
\end{center}

\end{document}